\documentclass[preprint,flushrt]{aastex}
\slugcomment{Submitted to \textit{The Astrophysical Journal}}
\usepackage{lineno}
\usepackage{url}
\linenumbers

\def\arcsec{\hbox{$^{\prime\prime}$}}
\def\arcmin{\hbox{$^{\prime}$}}
\def\deg{\hbox{$^\circ$}}
\def\Angst{$\buildrel _{\circ} \over {\mathrm{A}}$}
\def\Fermi{\textit{Fermi}}

\shorttitle{{\it Fermi}/LAT Observations of 4C\,+55.17}
\shortauthors{{\it Fermi}/LAT Collaboration}

\begin{document}

\title{{\it Fermi} Large Area Telescope Observations of the Active Galaxy 4C\,+55.17: Steady, Hard Gamma-Ray Emission and its Implications}
\author{
W.~McConville\altaffilmark{1,2,3}, 
L.~Ostorero\altaffilmark{4,5,6,7}
R.~Moderski\altaffilmark{8},
{\L}.~Stawarz\altaffilmark{9,10,11}
C.~C.~Cheung\altaffilmark{12,13}, 
M.~Ajello\altaffilmark{14}, 
A.~Bouvier\altaffilmark{15}
J.~Bregeon\altaffilmark{16}, 
D.~Donato\altaffilmark{17,2}, 
J.~Finke\altaffilmark{18}, 
A.~Furniss\altaffilmark{19}
J.~E.~McEnery\altaffilmark{1,2}, 
M.~E.~Monzani\altaffilmark{14}, 
M.~Orienti\altaffilmark{20,21},
L.~C.~Reyes\altaffilmark{22}, 
A.~Rossetti\altaffilmark{21}, 
D.~A.~Williams\altaffilmark{19}
}
\altaffiltext{1}{NASA Goddard Space Flight Center, Greenbelt, MD 20771, USA}
\altaffiltext{2}{Department of Physics and Department of Astronomy, University of Maryland, College Park, MD 20742, USA}
\altaffiltext{3}{email: wmcconvi@umd.edu}
\altaffiltext{4}{Dipartimento di Fisica Generale ``Amedeo Avogadro'', Universit\`a degli Studi di Torino, Via P. Giuria 1, I-10125 Torino, Italy}
\altaffiltext{5}{Istituto Nazionale di Fisica Nucleare (INFN), Sezione di Torino, Via P. Giuria 1, I-10125 Torino, Italy}
\altaffiltext{6}{Department of Physics \& Astronomy, University of Pennsylvania, 209 S. 33rd St., Philadelphia, PA 19104, USA}
\altaffiltext{7}{Harvard-Smithsonian Center for Astrophysics, 60 Garden Street, Cambridge, MA 02138, USA}
\altaffiltext{8}{Nicolaus Copernicus Astronomical Center, ul. Bartycka 18, 00-716 Warsaw, Poland}
\altaffiltext{9}{Institute of Space and Astronautical Science, JAXA, 3-1-1 Yoshinodai, Chuo-ku, Sagamihara, Kanagawa 252-5210, Japan}
\altaffiltext{10}{Astronomical Observatory, Jagiellonian University, 30-244 Krak\'ow, Poland}
\altaffiltext{11}{email: stawarz@astro.isas.jaxa.jp}
\altaffiltext{12}{National Research Council Research Associate, National Academy of Sciences, Washington, DC 20001, resident at Naval Research Laboratory, Washington, DC 20375, USA}
\altaffiltext{13}{email: Teddy.Cheung.ctr@nrl.navy.mil}
\altaffiltext{14}{W. W. Hansen Experimental Physics Laboratory, Kavli Institute for Particle Astrophysics and Cosmology, Department of Physics and SLAC National Accelerator Laboratory, Stanford University, Stanford, CA 94305, USA}
\altaffiltext{15}{Santa Cruz Institute for Particle Physics, Department of Physics and Department of Astronomy and Astrophysics, University of California at Santa Cruz, Santa Cruz, CA 95064, USA}
\altaffiltext{16}{Istituto Nazionale di Fisica Nucleare, Sezione di Pisa, I-56127 Pisa, Italy}
\altaffiltext{17}{Center for Research and Exploration in Space Science and Technology (CRESST) and NASA Goddard Space Flight Center, Greenbelt, MD 20771, USA}
\altaffiltext{18}{Space Science Division, Naval Research Laboratory, Washington, DC 20375, USA}
\altaffiltext{19}{Santa Cruz Institute for Particle Physics and Department of Physics, University of California, Santa Cruz, CA 95064, USA}
\altaffiltext{20}{Dipartimento di Astronomia, Universit\`a di Bologna, via Ranzani 1, I-40127, Bologna, Italy}
\altaffiltext{21}{Istituto di Radioastronomia – INAF, via Gobetti 101, 40129 Bologna, Italy}
\altaffiltext{22}{Kavli Institute for Cosmological Physics, University of Chicago, Chicago, IL 60637, USA}
\begin{abstract}

We report {\it Fermi}/LAT observations and broad-band spectral modeling of the
radio-loud active galaxy 4C\,+55.17 (z=0.896), formally classified as a
flat-spectrum radio quasar.  Using 19 months of all-sky survey {\it Fermi}/LAT
data, we detect a $\gamma$-ray continuum extending up to an observed energy of
$145$\ GeV, and furthermore we find no evidence of $\gamma$-ray variability in
the source over its observed history. We illustrate the implications of these
results in two different domains.  First, we investigate the origin of the
steady $\gamma$-ray emission, where we re-examine the common classification of
4C\,+55.17 as a quasar-hosted blazar and consider instead its possible nature
as a young radio source. We analyze and compare constraints on the source
physical parameters in both blazar and young radio source scenarios by means of
a detailed multiwavelength analysis and theoretical modeling of its broad-band
spectrum. Secondly, we show that the $\gamma$-ray spectrum may be formally
extrapolated into the very-high energy (VHE; $\geq 100$\,GeV) range at a flux
level detectable by the current generation of ground-based Cherenkov
telescopes. This enables us to place constraints on models of extragalactic
background light (EBL) within LAT energies and features the source as a
promising candidate for VHE studies of the Universe at an unprecedented
redshift of z=0.896.

\end{abstract}

\keywords{galaxies: active --- galaxies: individual (4C\,+55.17) --- galaxies: jets --- gamma rays: observations --- radiation mechanisms: non-thermal}

\section{Introduction}
\label{section-intro}

The radio-loud active galaxy 4C\,+55.17 (0954+556), formally classified as a
flat spectrum radio quasar (FSRQ), has a history of $\gamma$-ray observations
dating back to the EGRET era, as 3EG\,J0952+5501 \citep{har99,mat01} and
EGR\,J0957+5513 \citep{cas08}. Due to a relatively poor localization of the
EGRET source, however, the association of the $\gamma$-ray emitter with
4C\,+55.17 remained tentative at that time. After the successful launch of the
{\it Fermi} Gamma-Ray Space Telescope in June 2008, this association was on the
other hand quickly confirmed by the Large Area Telescope \citep[LAT;][]{atw09},
initially as 0FGL\,J0957.6+5522 \citep{bsl,lbas}, and most recently as
1FGL\,J0957.7+5523 \citep{1fgl}.  

The quasar classification of 4C\,+55.17 may be attributed to the presence of
broad optical emission lines in its spectrum \citep{wil95} and high optical/UV
core luminosity \citep[absolute $B$-band magnitude, $M_{\rm B} <
-23$;][]{ver06}.  Its redshift\footnote{Assuming a $\Lambda$CDM cosmology with
$H_{\rm 0}=71$\,km\,s$^{-1}$\,Mpc$^{-1}$, $\Omega_{\rm M}=0.27$, and
$\Omega_{\rm \Lambda}=0.73$, the luminosity distance $d_{\rm L}=5785$\,Mpc, and
the conversion scale is $1$\,mas\,$=7.8$\,pc.}, $z=0.896$, is based on the
detection of Ly$\alpha$ and CIV lines with the HST-FOS \citep{wil95} and Mg II
in the SDSS spectrum \citep{sch07}.  The optical-UV properties of the source,
together with its high $\gamma$-ray luminosity of the order $L_{\gamma} \simeq
10^{47}$\,erg\,s$^{-1}$, have in turn led to the common classification of
4C\,+55.17 as a blazar/FSRQ.  

However, 4C\,+55.17 also exhibits a number of morphological and spectral
properties that have placed its exact blazar/FSRQ classification into question
\citep{mar02,ros05}.  FSRQs are uniquely characterized by the presence of a
central compact radio core exhibiting a highly variable flat-spectrum
continuum, high brightness temperatures ($T_{\rm b}$), and, typically,
superluminal motions on VLBI scales \citep{urry95}.  Indeed, all of the
aforementioned radio properties are shared by the luminous blazars detected in
$\gamma$-rays: these are exclusively observed to possess compact, highly
polarized jets a few milli-arcseconds (mas) in angular size, and unresolved
radio cores with brightness temperatures in the range $T_{\rm
b}=10^{10}-10^{14}$\,K when observed at 5 GHz \citep{tay07} and 15 GHz
\citep{kov09}.  In comparison, 4C\,+55.17 demonstrates none of these
characteristics. To date, the source shows no evidence of blazar flaring at any
wavelength, nor any evidence of long-term variability, with the exception of a
$\sim 30\%$ optical flux-density change noted over a period of 7 years between
recent {\it Swift}/UVOT measurements and archival SDSS data (see
\S\,\ref{section-other_data} and \S\,\ref{section-model} for discussion).
Furthermore, the VLBI radio morphology of the source is extended over $\sim
400$\,pc (projected). The peak surface brightness in a VLBA 15 GHz image taken
from \citet{ros05} is found in the northernmost component and is clearly
resolved, with a corresponding brightness temperature $T_{\rm b} < 2 \times
10^{8}$ K \citep[consistent with a measurement at 5 GHz;][]{tay07}, which is
uncharacteristic of all the other known quasar-hosted $\gamma$-ray blazars.

\begin{figure}
\epsscale{0.95}
\centering
\plotone{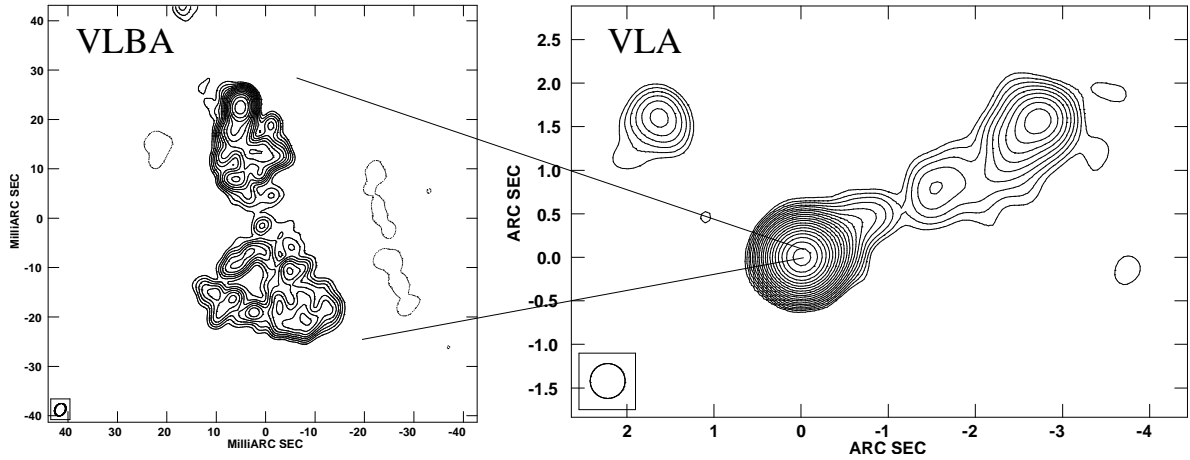}
 \caption{VLBA 5 GHz map (left) featuring the inner parsec-scale radio structure
 of 4C\,+55.17, reimaged using data from \citet{hel07}.  The beam size is 2.0
 mas $\times$ 1.6 mas (position angle = $-29.6\deg$), and the contour levels
 increase by factors of $\sqrt{2}$ beginning at 1 mJy/beam.  The resolved
 morphology has a total angular size of 53 mas (413 pc).  The VLA 5 GHz map
 (right) with a $0.4\arcsec$ beam (lowest contour is 2 mJy/beam increasing by
 factors of $\sqrt{2}$) shows the large scale radio structure
 \citep[from][]{tav07}.
}
\label{figure-radio}
\end{figure}

Based on the radio morphology of 4C\,+55.17, \cite{ros05} first suggested that
the source may in fact belong to the family of young radio sources
\citep[for a review, see][]{odea98}, rather than blazars. Such sources are
characterized by a very low radio variability (if any) and symmetric double
radio structures resembling ``classical doubles'' on much smaller scales:
linear sizes $\lesssim 1$\,kpc for compact symmetric objects (CSOs) and $\sim
1-15$\,kpc for medium symmetric objects \citep[MSOs;][]{aug06}, to be compared
with the typical linear sizes of ``regular'' Fanaroff-Riley type-II radio
galaxies of $\sim 100$\,kpc. In many cases, CSO sources are found to exhibit a
turnover in their radio spectra in the range of $0.5 - 10$\,GHz, as the
so-called “Gigahertz Peaked Spectrum (GPS) objects” do \citep{dev97};
similarly, MSO's often display turnover frequencies below $0.5$\,GHz, typical
of the Compact Steep Spectrum (CSS) class of sources \citep{fan90}.  The
overlap between CSO and GPS samples, as well as between samples of MSOs and CSS
sources, is however not complete \citep{sne00,aug06}.  In the case of
4C\,+55.17, the VLBI morphology at $5$\,GHz reveals two distinct emission
regions, to the north and south \citep[][see also
Figure\,\ref{figure-radio}]{ros05}, covering a total angular extent of
$53$\,mas ($=413$\,pc, projected). On the kpc scale, the source reaches
$4\arcsec.5$ ($\sim 35$\,kpc, projected), and it is resolved with the VLA in
three components, the central one hosting the VLBI structure.  The northern
component of the pc-scale emission features a compact region with a relatively
flat spectrum \citep[$\alpha=0.4; F_{\nu}\propto\nu^{-\alpha}$;][]{ros05},
which can be attributed to a core or a hotspot region, while the southern
component features a more diffuse and slightly steeper-spectrum ($\alpha=0.49$)
region.  \citet{ros05} have pointed out that these two components resemble more
compact hotspots and lobes, suggesting a CSO/MSO classification for this
object. The kpc-scale emission might thus be interpreted as a remnant of
previous jet activity, as this is a common feature among sources that show
evidence of intermittent behavior \citep[e.g.,][]{baum90,luo07,ori08}.  Under
the CSO/MSO framework, \citeauthor{ros05} found no core candidate between the
VLBA-scale lobes at a level $\gtrsim 2$\,mJy/beam in a $15$\,GHz map.

An 11-month comparison of the $\gamma$-ray variability and spectral properties
of 4C\,+55.17 against the other LAT FSRQs highlights the unusual nature of the
source \citep{1fgl,1lac}.  Among all of the sources originally detected in the
3-month LAT Bright AGN Sample \citep[LBAS;][]{lbas} that were classified as
FSRQs, 4C\,+55.17 is characterized by the lowest variability index
\citep{1fgl}.  In addition, the unusually hard $\gamma$-ray continuum (that is,
with a low photon index $\Gamma$) is found to be one of the hardest among
FSRQs in the 1st LAT AGN Catalog \citep[1LAC;][]{1lac}.  In fact, of those
sources included in the 1LAC (FSRQ or otherwise) with $>1$\,GeV flux greater
than or equal to that of 4C\,+55.17, only five -- all of which are BL Lac
objects (PKS~2155--304, Mkn 421, 3C 66A, PG~1553+113, and PKS~0447-439) --
appear with a harder $\gamma$-ray spectrum.

In this work, we re-examine the high-energy $\gamma$-ray ($>100$\,MeV)
properties of 4C\,+55.17 using 19 months of LAT all-sky survey data and discuss
the implications of these results in two domains.  First, we reconsider the
underlying physical processes responsible for the $\gamma$-ray emission through
detailed broadband modeling of the source in the context of two scenarios:
``young radio source'' and ``blazar.''  In addition, we demonstrate that the
unusual properties of the source make it an ideal candidate for studying the
high-redshift universe at very-high energies (VHE), in particular for placing
constraints on the level of extragalactic background light (EBL).  The paper is
organized as follows.  Section\,\ref{section-observations} details the analysis
of 19 months of LAT data and discusses the supporting multiwavelength
observations.  In particular, section\,\ref{section-lat} focuses on the LAT
data reduction, presenting new spatial (localization), spectral, and
variability analysis, including a detailed analysis of the 145 GeV photon
detection associated with the source (see also Appendix\ \ref{section-photon}).
Section \ref{section-other_data} discusses the multiwavelength observations,
including analysis of archival radio and {\it Swift} X-ray and optical data, as
well as a new hard X-ray detection with the {\it Swift} Burst Alert Telescope
(BAT).  Spectral properties and classification of 4C\,+55.17 are discussed in
section\,\ref{section-discussion}. We follow with a detailed analysis of the
high energy spectrum of 4C\,+55.17, where we place constraints on models of EBL
and discuss the implications of the 145 GeV photon detection to future VHE
observations of the source (\ \S\,\ref{section-EBL}).  Our conclusions are
presented in section\,\ref{section-conclusions}.

\section{Observations}
\label{section-observations}

\subsection{{\it Fermi}/LAT Observations}
\label{section-lat}

The {\it Fermi}/LAT is a pair creation telescope designed to cover the energy
range from $\sim 20$\,MeV to $>300$\,GeV \citep{atw09}. The LAT instrument
features an improved angular resolution ($\theta_{\rm 68\%}=0.8\deg$ at
$1$\,GeV) over previous instruments and a large field-of-view of $2.4$\,sr. The
nominal mode of operation is an all-sky survey mode, which provides nearly
uniform sky coverage approximately every 3 hours. The following analysis is
comprised of 19 months of nominal all-sky survey data extracted from a $10\deg$
region of interest (ROI) around the J2000.0 radio position of 4C\,+55.17
\citep[R.A.\,$=09^{\rm h}57^{\rm m}38.1844^{\rm s}$, Decl.\,$=55\deg
22\arcmin57.769\arcsec$;][]{fey04} and covers the mission elapsed time (MET)
$239557417$ to $289440000$ (August 4, 2008 through March 4, 2010). A 100-second
interval at MET $251059717$ was removed in order to avoid contamination from
GRB\,081215A, which fell within the ROI.  Event selections include the
``diffuse'' event class \citep{atw09} recommended for point source analysis, a
zenith angle cut of $<105\deg$ to avoid contamination from the earth limb, and
rocking angle cuts at $43\deg$ and $52\deg$, respectively, for times
corresponding to a change in the instrument's rocking angle from $39\deg$ to
$50\deg$ that occured on September 3, 2009 (MET $273628805$). Science Tools
\texttt{v9r16p1} and instrument response functions (IRFs)
\texttt{P6\_V3\_DIFFUSE} were used for this analysis.

The 19 month LAT localization of 4C\,+55.17 was determined using
\texttt{gtfindsrc}, resulting in a best-fit position (J2000.0) of R.A.\,$=
09^{\rm h}57^{\rm m}40^{\rm s}$, Decl.\,$ = 55\deg23\arcmin40\arcsec$, which is
$0.012\deg=0.7\arcmin$ offset from the radio position and falls within the
$95\%$ error circle $r_{95\%}=0.017\deg=1.0\arcmin$ (statistical only).  In
order to model the $\gamma$-ray emission, all point sources from the 1FGL
catalog \citep{1fgl} within $15\deg$ of the source were included. Sources
within $10\deg$ of the 4C\,+55.17 radio position were modeled with their flux
and spectral parameters set free, while those sources that fell outside the
$10\deg$ ROI were fixed at their catalog values. The diffuse background was
modelled using the
recommended\footnote{\url{http://fermi.gsfc.nasa.gov/ssc/data/access/lat/BackgroundModels.html}}
Galactic diffuse \texttt{gll\_iem\_v02.fit} along with the corresponding
isotropic spectral template \texttt{isotropic\_iem\_v02.txt}.

Prior to fitting the spectrum, the high energy photons attributable to the
4C\,+55.17 position (both radio and $\gamma$-ray) were found by comparing the
energy and incoming angle $\theta$ (defined with respect to the spacecraft
zenith) of each photon within the ROI to the 95\% containment radius of the
point spread function defined by the \texttt{P6\_V3\_DIFFUSE} IRFs. Included
among those photons was a $145$\,GeV event at an angular separation of
$0.06\deg$ (R.A.\,$=09^{\rm h}58^{\rm m}03^{\rm s}$, Decl.\,=$55\deg 24\arcmin
00\arcsec$) from the 4C\,+55.17 position, falling well within the ${95\%}$
containment radius for the given energy and angle of incidence.  Through an
analysis of the event diagnostics, the photon nature of this event is confirmed
here for the first time (for further details regarding the 145\,GeV event
analysis, see Appendix\ \ref{section-photon}).  In addition, several photons in
the $\sim 30 - 55$ GeV range were also detected.  The association of the
145\,GeV photon with 4C\,+55.17 tentatively places it as the highest-redshift
source to be observed at VHE to date.  

A spectral analysis of 4C\,+55.17 was performed with \texttt{gtlike} using the
LAT data between $100$\,MeV and $300$\,GeV. Spectral data points were first
obtained by fitting each of 9 equal logarithmically spaced energy bins to a
separate power law with index and prefactor parameters set free. From the
resulting data points, a break in the spectrum could be seen to occur at $\sim
1.6$\,GeV. This was confirmed by performing an independent unbinned likelihood
fit over all the data from $100$\,MeV to $20$\,GeV using power law, log
parabola, and broken power law models, with the break energy of the broken
power law fixed at the peak in the $\nu F_{\nu}$ representation ($E_{\rm br}
\sim 1.6$\,GeV). The maximum energy of 20 GeV was chosen in order to avoid
fitting any portion of the spectrum that may be significantly attenuated by the
EBL.  A likelihood ratio test \citep{mat96} resulted in a $4.1\sigma$
improvement of the broken power law over the single power law used in previous
analyses of the source \citep{bsl,1fgl}, as compared to a $3.8\sigma$
improvement over the power law from the log parabola.  We therefore consider
the broken power law to be the most accurate representation of the intrinsic
$\gamma$-ray spectrum of the source.  

\begin{figure}
\epsscale{1.0}
\centering
\plotone{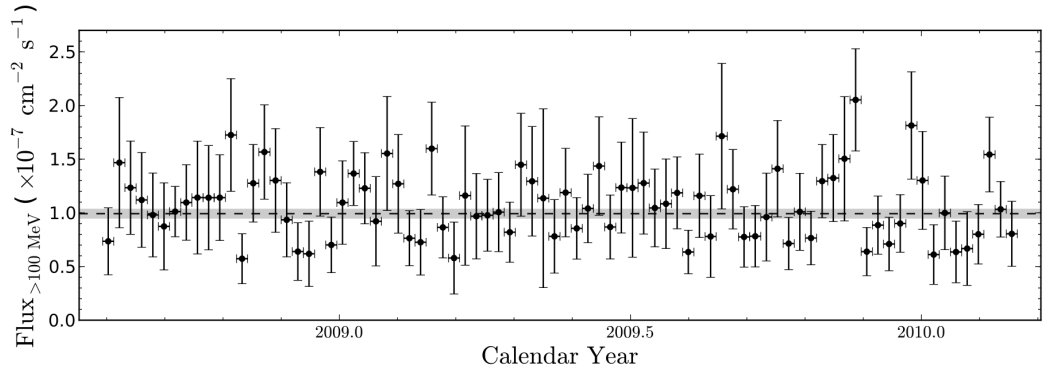}
 \caption{\Fermi/LAT 19-month $\gamma$-ray light curve of 4C\,+55.17 divided into 7 day bins.  
All points represent $>3\sigma$ detections and are plotted along with their
statistical errors.  The dashed horizontal line and gray region represent the
weighted mean and corresponding error derived from all $>3\sigma$ detections
over the observing period. 
}
\label{figure-lc}
\end{figure}

To test the $\gamma$-ray variability over the 19-month period, we made light
curves in time bins of 7 and 28 days. Due to the limited statistics over each
interval, the source was fit to a single power-law in each bin, with index and
prefactor parameters free. To improve the fit convergence, point sources in the
ROI were included only if they were detected with a test statistic
\citep[TS;][]{mat96} greater than 1 ($\sim 1\sigma$).  The resulting light
curve ($>100$\,MeV), divided into 7 day bins, is shown in
Figure\,\ref{figure-lc}.  The variability of 4C\,+55.17 was analyzed by means
of a $\chi^{2}$ test, where we assumed the model describing the data to be a
constant straight line with intercept equal to the weighted mean of all
$>3\sigma$ detections. This test yielded a $\chi^{2}$ probability $P(\chi^{2}
\ge \chi^{2}_{\rm obs})$ of 0.96 and 0.87 for the 7 day and 28 day light
curves, respectively, and was thus in agreement with the tested hypothesis.  We
therefore found no evidence of variability in $\gamma$-rays over the 19-month
LAT observing period, consistent with the previous 11-month lightcurve analysis
($\sim\,30$ day bins) from \citet{1fgl}. In addition, the weighted mean for
this period was found to be $(9.5 \pm 0.4_{\rm stat}+0.83_{\rm sys}-0.49_{\rm
sys}) \times 10^{-8}$\,ph\,cm$^{-2}$\,s$^{-1}$, which is consistent with the
EGRET measured flux of $(9.1 \pm 1.6) \times 10^{-8}$\,ph\,cm$^{-2}$\,s$^{-1}$
\citep{har99} as well. Systematic uncertainties on the LAT flux were determined
by bracketing the instrument effective area to values of 10\%, 5\%, and 20\%
their nominal values at $\log (\rm E/MeV)=2, 2.75,$ and $4$, respectively.
We note that these findings differ from those of \cite{ner10}, who claim
variability between the EGRET and LAT measured fluxes.  We believe this
discrepancy lies in a mis-quoted value of the EGRET flux. 

\subsection{Multiwavelength Data}
\label{section-other_data}

\subsubsection{X-ray}
\label{section-x-ray}

We analyzed all {\it Swift} \citep{geh04} data obtained over the 19-month LAT
observing period, which consisted of three X-ray Telescope
\citep[XRT;][]{bur05} snapshots (1.6-4.5~ks), in order to check the X-ray state
of the source. We used the xrtgrblc script (available in the HEASoft package
version 6.8) to analyze the XRT observations: we reprocessed the data stored in
the HEASARC archive using the latest XRT calibration database (20091130),
selecting the events with 0-12 grades in photon counting mode (PC). The scripts
chose the optimal source and background extraction regions based on the source
intensity: the X-ray photons were extracted using a $25\arcsec$ circle for the
source and an annulus with $50\arcsec-150\arcsec$ inner-outer radius for the
background.  Adding all of the exposure and performing a C-statistic fit from
$0.3-10$\,keV using XSpec12, we found the best fit obtained to be a power law
with absorption fixed at the galactic value ($N_{\rm
H}=9\times10^{19}$\,cm$^{-2}$), where we obtained the photon index $\Gamma=1.84
\pm 0.19$, with an absorbed flux of
$(8.3^{+1.7}_{-1.4})\times10^{-13}$\,erg\,cm$^{-2}$\,s$^{-1}$ and an unabsorbed
flux of $(8.5^{+1.7}_{-1.4})\times10^{-13}$\,erg\,cm$^{-2}$\,s$^{-1}$.
Comparing each of the individual observations, no X-ray variability was found,
with all measurements falling within the joint errors. These results were also
compared with previous {\it Chandra} data \citep{tav07} obtained June 16, 2004,
where the flux was found again to be non-variable within the statistical
errors.  Finally, historical X-ray data from ROSAT \citep{com97} obtained
November 7, 1993 were included in the spectral energy distribution (SED)
modeling to further constrain the soft X-ray portion of the spectrum.

In the hard X-rays, data from the {\it Swift}/BAT \citep{aje08,aje09} were
analyzed using five years of cumulative exposure from November\ $2005-2010$.
We detect the source for the first time in the hard X-ray band, with a
$15-150$\,keV flux of $(6.75^{+0.38}_{-5.21}) \times
10^{-12}$\,erg\,cm$^{-2}$\,s$^{-1}$ and a power-law photon index,
$\Gamma=1.79^{+1.17}_{-0.84}$.

\subsubsection{Optical \& Infrared}

During each of the three {\it Swift} pointings in 2009, Ultra-Violet/Optical
Telescope \citep[UVOT;][]{rom05} observations were also obtained.  Data were
obtained in all 6 filters in the first two epochs, and the last epoch with only
the $W2$ filter. The data reduction and analysis was performed using the
uvotgrblc script, which reprocesses the data stored in the HEASARC using the
latest UVOT calibration database (20100129). The optimal source and background
extraction regions were a $5\arcsec$ circle and a $27\arcsec-35\arcsec$
annulus, respectively. Table\,\ref{tab:uvot} summarizes these observations. A
comparison of the results between each epoch shows the source to fall within
the joint errors in flux in the optical to UV bands across all three epochs.
These results were also compared with archival SDSS data from February 2, 2002
\citep{ade08}. A comparison of the UVOT and SDSS $U$-band flux densities shows
an increase from $(0.187 \pm 0.003)$\,mJy in the SDSS data to $(0.250 \pm
0.007)$\,mJy in the UVOT data, indicating a $\sim30\%$ rise in flux over 7
years.  In addition, the UVOT $V$- and $B$-band flux densities were averaged
using a least-squares approach to a linear fit and compared with the SDSS
$g$-band, which fell between the two.  The average of the UVOT $V$- and
$B$-bands, measured at $(0.305 \pm 0.014)$\,mJy, shows a similar $\sim 25\%$
increase from the SDSS measured value of $(0.240 \pm 0.011)$\,mJy.  A
comparison of the {\it Swift} UVOT measurements to the continuum flux
underlying the Ly$\alpha$ line obtained by HST-FOS in 1993 \citep{wil95} shows
the fluxes to be equal between these two periods.

In the near-infrared, we included historical data from the 2MASS Point Source
Catalog \citep{cut03}, for which the absolute calibration was taken from
\citet{coh03}. All infrared, optical, and ultraviolet data were dereddened by
means of the extinction laws given by \citet{car89}, assuming a $B$-band
Galactic extinction ($A_{\mathrm B}=0.038$) as determined via \citet{sch98},
and a ratio of total to selective absorption at $V$ equal to $R_{\mathrm
V}=3.09$ \citep{rie85}.

\subsubsection{Radio}

\begin{figure}
\epsscale{0.90}
\plotone{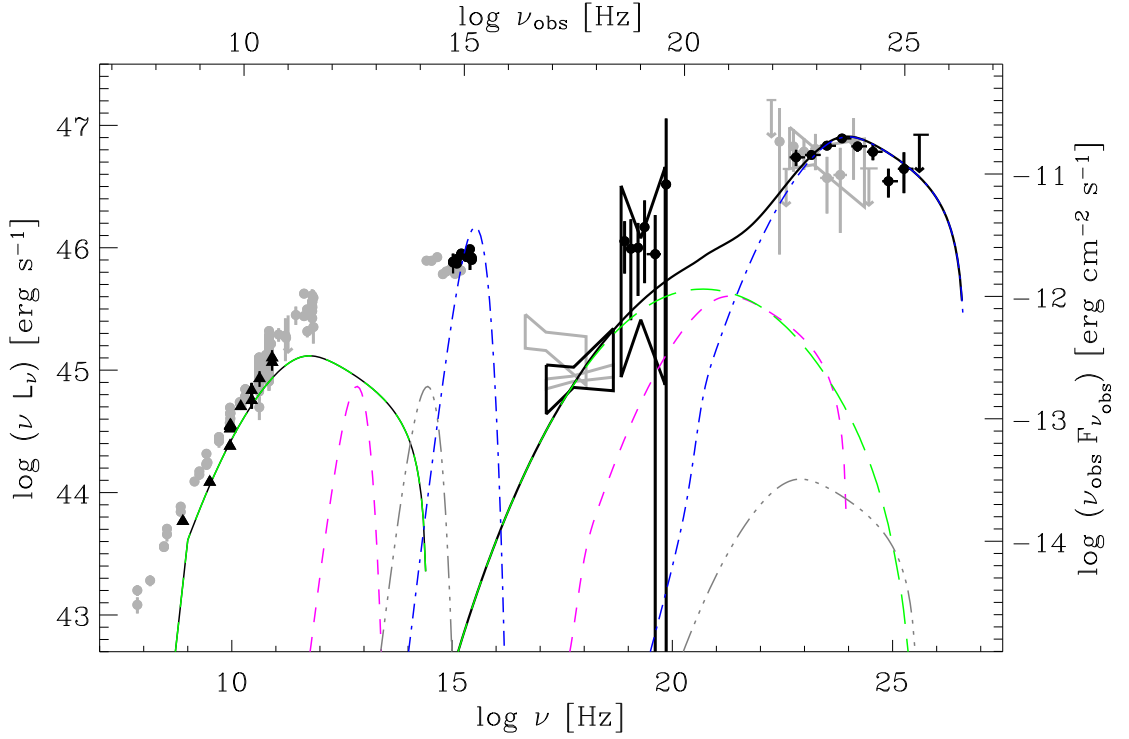}
 \caption{The CSO model of 4C\,+55.17 versus multiwavelength data, including the new 
LAT spectrum along with contemporaneous data with {\it Swift} XRT, BAT, and
UVOT (black bullets).  Archival detections (gray) with EGRET \citep{har99},
ROSAT, Chandra, SDSS, 2MASS, 5-year integrated WMAP, and historic radio data
are also included, as well as archival VLA measurements (black triangles) of
the inner $\sim 400$pc radio structure (see \S\,\ref{section-other_data}).
De-absorption of the observed {\it Fermi} spectral points using the
\citet{fin10} EBL model was applied in order to properly model the intrinsic
$\gamma$-ray spectrum.  Black curves indicate the total non-thermal emission of
the lobes, with the long-dashed/green representing the contribution from
synchrotron self-Compton (SSC).  Dashed/pink, dash-dot-dotted/gray, and
dash-dotted/blue blackbody-type peaks represent the dusty torus, starlight, and
the UV disk emission components, respectively, along with their corresponding
inverse-Compton components as required by the model.  
}
\label{figure-model}
\end{figure}

\begin{figure}
\epsscale{0.90}
\plotone{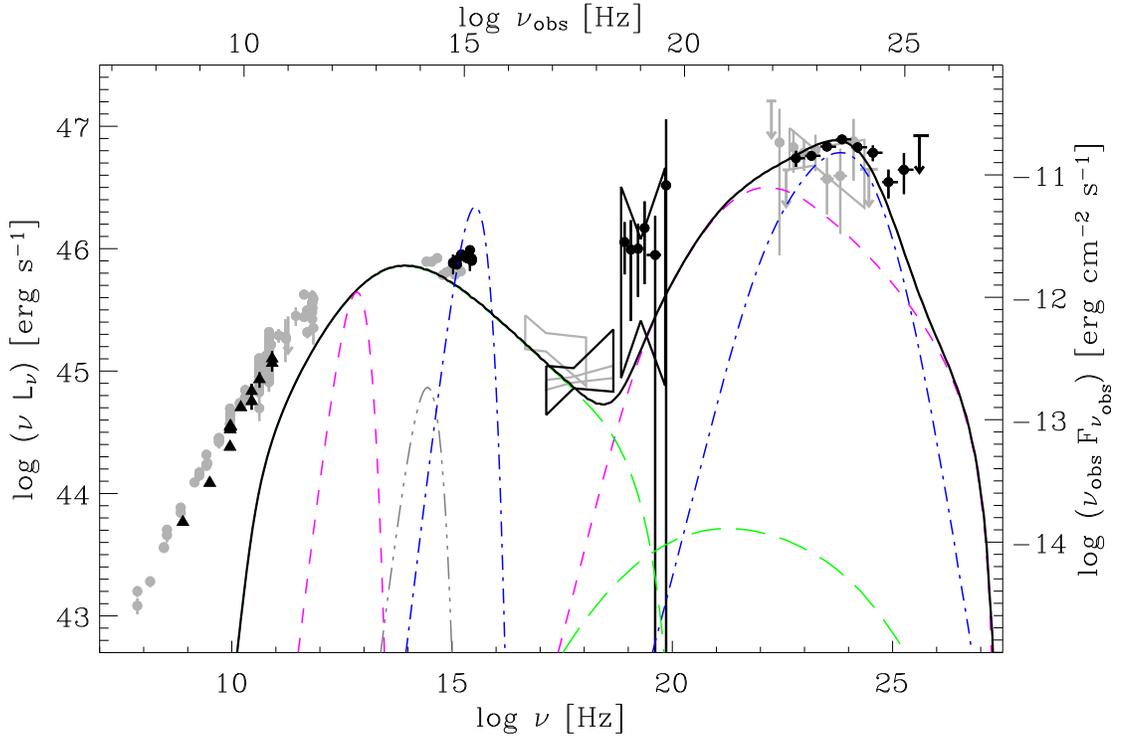}
 \caption{Blazar fit using multi-wavelength data for 4C\,+55.17.  
Indicated are the individual contributions from synchrotron and SSC
(long-dashed/green), as well as IC scattering off of the reprocessed UV disk
emission from the broad line region (dash-dotted/blue), dusty torus
(dashed/pink), and host galaxy (dash-dot-dotted/gray); the black curve
indicates the total of these components. As in Fig.\,\ref{figure-model}, the
dashed/pink, dash-dot-dotted/gray, and dash-dotted/blue blackbody-type peaks
represent the dusty torus, starlight, and the UV disk emission components,
respectively, along with their corresponding inverse-Compton components as
required by the model.
}
\label{figure-model-bl}
\end{figure}

To model the $\gamma$-ray emission in 4C+55.17 (sec.~3.1), we compiled
integrated radio to sub-mm measurements of the source
\citep{blo94,hua98,rei98,jen10}, including 5-year WMAP data \citep{wri09}, and
other archival data from the NASA/IPAC Extragalactic Database (NED). In order
to isolate the total radio flux from the inner $\sim400$ pc scale
structure\footnote{The kpc-scale radio emission is not expected to contribute
significantly toward the modeling of the high energy portion of the spectrum
(see\,\S\ref{section-cso}\,\&\,\S\ref{section-blazar}).}, we re-analyzed
several archival VLA data sets from 5 to 43 GHz (see Figures \ref{figure-model}
and \ref{figure-model-bl}). The typical resolutions are $\sim 0.1\arcsec$ to
$0.4\arcsec$, ensuring a total measurement of the $\sim50$ mas scale structure
without loss of flux as in the VLBI observations \citep[e.g.,][]{ros05}.  We
also include similar measurements from previously published VLA 5 and 8.4 GHz
\citep{rei95,mye03,tav07} and MERLIN 0.4 and 1.7 GHz \citep{rei95} maps.

The radio variability properties of 4C\,+55.17 are important for assessing its
nature. We therefore searched the literature for various archival radio to
sub-mm monitoring observations of the source
\citep[e.g.][]{alt76,war81,sei83,jen10}, including $22$ and $37$ GHz data from
the Mets\"{a}hovi monitoring program \citep{ter98,ter04,ter05}. While the
\citet{war81} data was not directly available, we note from the literature that
the authors found the source to be non-variable.  Variability in each of the
remaining cases was measured by applying a statistical $\chi^{2}$ test of the
available data using the hypothesis of a constant source with flux equal to the
weighted mean.  The results were consistent with the tested hypothesis in each
case, with the exception of the Mets\"{a}hovi data, which yielded probabilities
P($\chi^{2} \ge \chi^{2}_{\rm obs}$) of $6.44\times 10^{-56}$ and
$8.56\times10^{-24}$ at $22$ GHz and $37$ GHz, respectively. To quantify this
variability, we compared fractional variability indices using the formula $\rm
Var_{\Delta S}=(S_{\rm max}-S_{\rm min})/S_{\rm min}$ used in a variability
study of GPS sources \citep{tor07}, where we obtained values of $3.5$ and
$1.43$ at $22$ and $37$ GHz, respectively.  The $22$ GHz value fell slightly
above the nominal variability threshold of 3.0 set by \citet{tor07} as an upper
limit for the bona fide GPS sources.  This result, however, arose due to a
single outlying flux measurement at 22 GHz of $0.32\pm0.09$\,Jy which occurred
$\sim40$ minutes after a previous measurement of $1.12\pm0.08$\,Jy at the same
frequency\footnote{Variability within hour timescales is rare at the
frequencies observed by Mets\"{a}hovi (A.  Lahteenmaki, T.  Hovatta, \& M.
Tornikoski, private communication 2010)}.  Removing this questionable flux
point and performing the test again resulted in a fractional variability index
of $0.89$, which fell well within the proposed threshold for genuine GPS
galaxies.  We therefore find the degree of variability in 4C\,+55.17 to be
consistent with the behavior of confirmed young radio galaxies, rather than
blazars.

\section{Results}
\label{section-discussion}

\subsection{Modeling \& Classification}
\label{section-model}
\subsubsection{CSO Modeling}
\label{section-cso}

As noted in the introduction, there are several reasons to consider the
possible nature of 4C\,+55.17 as an example of a luminous AGN exhibiting
recurrent jet activity, with young and
symmetric (CSO-type) inner radio structure instead of a ``core-jet'' morphology
typical of blazars.  While the physical nature and the origin of the CSOs is at
some level still debated, the most likely and widely accepted hypothesis states
that they are the young versions of present-day extended radio galaxies
\citep{phil82,fan95}.  In the alternative explanation, these sources are
considered to be of a similar age to normal radio galaxies, but only
confined/frustrated due to dramatic interactions with a surrounding dense gas
in their host galaxies \citep{bre84,wil84}. The latter scenario is however
inconsistent with the lack of observational evidence for the amount of ambient
gas required to supply sufficient confinement {\bibpunct[; ]{(}{)}{;}{a}{}{,}
\citep[][see, however, \citealp{gar07} for notable
exceptions]{young93,car94,car98,sie05}\bibpunct[, ]{(}{)}{;}{a}{}{,}}. More
promising is therefore the ``youth'' scenario for CSOs, for which a number of
evolutionary models were proposed \citep{beg96,young97,per02,kaw06}.

While many observational properties of 4C\,+55.17 make its classification as a
young radio source compelling, it is also worth noting the characteristics that
could make such a classification potentially difficult.  For example, if
4C\,+55.17 is indeed a CSO, it is the only such object to be identified as a
$\gamma$-ray emitter in 1FGL/1LAC, with a GeV flux nearly an order of magnitude
higher than the lower limit of the complete flux-limited subsample within the
1LAC catalog \citep{1lac}. This would immediately set the object apart as an
outstanding member of its class.  In addition, the relatively high radio
polarization of the source \citep[$\sim$\,$3\%$ in a $\sim$\,$0.2\arcsec$
resolution VLA $8.4$\,GHz image;][]{jac07}, is uncharacteristic of the
typically low ($<1\%$) radio polarization seen among CSOs \citep{rea96},
although polarized emission from CSOs has occasionally been found
\citep[e.g.,][]{gug07}. The low polarization of CSOs, which are entirely
embedded within the inner regions of the host galaxy, is often attributed to
the large expected Faraday depths of the surrounding interstellar medium
\citep{burn66,bic97,gug07}. The surrounding medium may also play a key role in
shaping the spectral turnover seen in the GPS class of young radio sources,
through the free-free absorption (FFA) process \citep[either internal or
external to the emission region;][]{bic97,beg99,pec99}. The nature of the
absorber is however still widely debated, and both the synchrotron
self-absorption (SSA) and FFA processes are considered as viable options
\citep{odea97,sne00}.

If FFA effects are indeed responsible for the spectral turnover in GPS sources,
then the relatively flat ($\alpha \simeq 0.4-0.5$) power-law radio continuum of
4C\,+55.17, which shows no indication of a low-energy turnover, may indicate an
exceptionally small amount of ionized ambient gas in the vicinity of its young
radio structure. More specifically, if the radio absorber may be identified
with ionization-bounded hydrogen clouds of interstellar matter present at pc to
kpc distances from the center and engulfed by the expanding lobes, as proposed
by \citet{beg99} and advocated by \citet{sta08}, and if a significant part of
this gas has been evacuated prior the onset of new jet activity, then one would
expect much less severe absorption of the low frequency radio emission,
resulting in a lower turnover frequency compared to that of GPS galaxies.  In
this case, the relatively high polarization of 4C\,+55.17 (as for a young radio
source) would find a natural and straightforward explanation as well.

In considering the hypothesis outlined above, and in order to investigate the
$\gamma$-ray emission detected from 4C\,+55.17 in a framework that is more
consistent with the observed properties of the source, we apply the dynamical
model for the broad-band emission of CSOs proposed by \citet{sta08} and
successfully tested against a sample of X-ray detected young radio galaxies of
the CSO type by \citet{ost10}. In this model, the newly born relativistic jets
propagate across the inner region of the host galaxy and inject
ultrarelativistic electrons into the compact lobes. These electrons, which
provide the bulk of the internal lobes' pressure, cool radiatively and
adiabatically within the sub-relativisticly expanding plasma, thus producing
isotropic synchrotron (radio) and IC (X-ray to $\gamma$-ray) radiation. In the
model, the broad-band emission spectra are evaluated self-consistently for a
given set of the initial parameters of the central engine and of the host
galaxy, taking into account the time-dependent evolution of the radiating
electrons. For a given linear size of the system, which is uniquely related to
a particular age of the system, the observed broad-band emission spectrum is
given as a snapshot of the evolving multiwavelength radiation of the lobes.
Based on this model, \citet{sta08} argued that, in fact, young radio galaxies
should be detected by {\it Fermi}/LAT at GeV photon energies, albeit at low
flux levels and after an exposure longer than one year.  Other (physically
distinct) scenarios for the production of soft, high energy, and VHE
$\gamma$-rays in the lobes and hotspots of young radio galaxies have been
proposed and investigated by \citet{kino07,kino09} and \citet{kino11}.

In the more detailed description of the model, the jets with total kinetic
power ($L_{\rm j}$) propagate with the advance velocity ($v_{\rm h}$) in the
interstellar medium, characterized by a given number density ($n_{\rm ext}$).
At a particular instant of the source evolution, the inflated lobes will have a
corresponding linear size ($LS$). The electrons injected through the
termination shock into the lobes with the intrinsically broken power-law energy
distribution cool due to the synchrotron and IC processes.  The most relevant
ambient photon fields for the IC scattering are the UV emission of the
accretion disk (mean photon energy $\varepsilon_{\rm disk} = 10$\,eV, disk
luminosity $L_{\rm disk}$), the starlight ($\varepsilon_{\rm star} = 0.83$\,eV,
host luminosity $L_{\rm star}$), and the infrared emission of the obscuring
nuclear torus ($\varepsilon_{\rm dust} = 0.02$\,eV, dust luminosity $L_{\rm
dust}$). The magnetic field intensity is expressed in terms of the ratio of
energy densities stored in the radiating electrons and the magnetic field,
$U_{\rm e}/U_{\rm B}$, which is constant during the expansion of the radiating
plasma. Note, however, that $U_{\rm e}$ and $U_{\rm B}$, as well as the energy
densities of the ambient photon fields (and hence the electron cooling
conditions) do change with time, and therefore depend on $LS$ \citep[see][for
more details]{sta08}.

The fit of the ``young radio source'' model to the collected broad-band dataset
for 4C\,+55.17 is illustrated in Figure\,\ref{figure-model}.  In fitting the
SED, we assume that the projected source size of the inner radio structure ($LS
\simeq 400$\,pc) is equal to the actual source size, though we note that this
may be underestimated due to possible projection effects. Indeed, some amount
of projection off the plane of the sky is required to account for the presence
of the intense disk-related optical/UV continuum and the broad optical emission
lines in the spectrum of 4C\,+55.17 (which might otherwise be completely
obscured), as well as to account for the asymmetry in brightness between the
two lobes.  In fitting the broadband SED, the following model-free parameters
were obtained: $L_{\rm j} \simeq 6.6 \times 10^{47}$\,erg\,s$^{-1}$, $L_{\rm
disk} \simeq 2 \times 10^{46}$\,erg\,s$^{-1}$, $L_{\rm star} \simeq
10^{45}$\,erg\,s$^{-1}$, $L_{\rm dust} \simeq 10^{45}$\,erg\,s$^{-1}$, $U_{\rm
e}/U_{\rm B} \simeq 160$, $v_{\rm h} \simeq 0.3 c$, and $n_{\rm ext} \simeq 
0.1$\,cm$^{-3}$. The injection electron energy distribution is characterized by
the minimum, break, and maximum electron Lorentz factors, $\gamma_{\rm min}
\simeq 1$, $\gamma_{\rm br} \simeq 2 \times 10^4$, and $\gamma_{\rm max} \simeq
4 \times 10^5$, respectively, as well as by the low- and high-energy electron
spectral indices, $s_1 \simeq 0.5$ and $s_2 \simeq 2.5$. The model fits quite
successfully all the relevant data points within the low-frequency (radio) and
high-frequency (hard X-ray to $\gamma$-ray IC component) ranges; it also
reproduces nicely the spectral break within the {\it Fermi}/LAT photon energy
range.  We note that in our modeling here and below we do not consider
$\gamma$-ray absorption effects related to the direct or reprocessed emission
of the accretion disk, which may lead to the attenuation of the lobes' (or
jets') emission at photon energies $> 100$\,GeV \citep[see in this
context][]{tanaka11}.

Looking closely at the UV part of the spectrum, we note an approximate factor
of two difference between what is observed and what is required for producing
the appropriate luminosity in IC-scattered $\gamma$ rays. This can be resolved
by recalling that in the framework of the model the optical/UV photon energy
range is dominated by the thermal UV disk emission that may suffer from some
non-neglible obscuration by the circumnuclear dust for moderate inclinations of
the source to the line of sight.  Also worth noting are the variation
timescales of the disk, which are governed by the viscous motion within tens of
gravitational radii from the black hole \citep{col01}.  This can account for
the $\sim 30 \%$ variation over seven years seen between the optical
measurements from UVOT and SDSS (see \S\,\ref{section-other_data}). On the
other hand, the CSO-related non-thermal IC emission is expected to be
non-variable in accordance with the observations, because this emission is
produced within the hundred-pc-scale and sub-relativistictically expanding
lobes, and hence the UV photons seen by the lobes' electrons will be averaged
over the entire spatial extent of the radio structure. Here we do not model the
accretion-related emission in detail, but only roughly represent it as a
blackbody component for the purpose of the evaluation of the IC radiation of
the lobes. Likewise, the steep-spectrum soft X-ray continuum is not accounted
for by the IC emission of compact lobes and instead may be attributed to the
radiative output of the accretion disk and its corona \citep[see][for the X-ray
properties of young radio sources]{sie08,sie09}. Yet it should be also noted
that the particular CSO model presented here cannot account for the
millimeter--to--near infrared emission of 4C\,+55.17. In the framework of the
discussed scenario, this has to be attributed to the radiation of the
underlying jet, and not of the compact lobes.

The physical parameters of 4C\,+55.17 emerging from the model fit presented
above may be compared with the physical parameters of bona fide young radio
galaxies derived in the framework of the same model by \citet{ost10}.  The most
significant differences can be noted in the kinetic luminosity of the jet
($L_{\rm j}$), the UV luminosity of the accretion disk ($L_{\rm disk}$), and
the electron--to--magnetic field energy density ratio ($U_{\rm e}/U_{\rm B}$).
In particular, the jet and the disk luminosities of 4C\,+55.17 are higher (by
one to two orders of magnitude, on average) than the analogous luminosities of
GPS radio galaxies. This is in fact expected, since the analyzed source is much
more powerful than the relatively low-power radio galaxies modeled by
\citet{ost10}. The disk luminosity obtained from the fit can also be compared
with the expected value based on the total luminosity of emission in broad
lines ($L_{\rm BLR}$).  Using eq. (1) in \citet{cel97}, along with the line
fluxes of 4C\,+55.17 obtained in \citet{wil95} and the line ratios from
\citet{fra91}, we estimate the value of $L_{\rm BLR}$ to be $1.2 \times
10^{45}$\,erg\,s$^{-1}$.  Using the approximation $L_{\rm disk} \simeq 10
\times L_{\rm BLR}$, we thus obtain $L_{\rm disk} \simeq 1.2 \times
10^{46}$\,erg\,s$^{-1}$, which again falls within a factor of two of the value
obtained through the model, consistent with the level of uncertainty expected
using this method.

\subsubsection{Blazar Modeling}
\label{section-blazar}

As already noted in the introduction, the lack of pronounced variability and
resolved VLBI structure in 4C\,+55.17 would make it a highly unusual case of a
blazar/FSRQ.  Still, it is a worthwhile exercise to consider the physical
parameters implied from the blazar model. In the framework of the blazar
scenario the observed non-thermal emission of this source, including the
$\gamma$-ray flux detected by {\it Fermi}/LAT, is expected to originate in the
innermost parts of a relativistic jet that is closely aligned with the line of
sight \citep[e.g.,][]{sik94}. In this case, the broad-band emission of
4C\,+55.17 should be strongly Doppler boosted in the observer rest frame, and
variable on short (days to weeks) timescales. The expected size of the blazar
emission region (sub-pc), which is orders of magnitude smaller than the linear
size of the resolved inner radio structure discussed previously ($\sim
400$\,pc), as well as the presence of relativistic beaming effects, constitute
the main differences between the ``blazar'' and ``young radio source''
scenarios. 

In order to model the broad-band spectrum of 4C\,+55.17 as a blazar emission,
we apply the dynamical model BLAZAR developed by \citet{mod03} and later
updated by \citet{mod05} for the correct treatment of the Klein-Nishina regime
\citep[for applications of the model, see e.g.][]{sik08,kat08}.  The model
describes the production of the non-thermal emission by ultrarelativistic
electrons, which are accelerated in situ within thin shells of plasma
propagating along a conical relativistic jet (bulk Lorentz factor, $\Gamma_{\rm
j} \gg 1$, jet opening angle $\theta_{\rm j} \sim 1 / \Gamma_{\rm j}$) and
which carry a fraction $L_{\rm e} / L_{\rm j}$ of the jet kinetic power. The
acceleration process is attributed to the Fermi mechanism operating at strong
shocks that are formed within the outflow as a result of the shells'
collisions, which take place at distances greater than $r_0$ from the jet base,
resulting in the injection of a broken power-law electron energy distribution
into an emission region of linear size $R$ and magnetic field intensity $B$.
The non-thermal emission evaluated at $r \simeq R / \theta_{\rm j} \gtrsim
r_{\rm 0}$ includes the synchrotron and IC components, with the target photons
for the inverse-Compton scattering provided by the jet synchrotron radiation
and the external photon fields (predominantly accretion disk emission
reprocessed in the broad line region and within the dusty torus).

The BLAZAR fit to the broad-band spectrum of 4C\,+55.17 is shown in
Figure~\ref{figure-model-bl}.  The fit was obtained with the following free
parameters of the model: $L_{\rm j} \geq L_{\rm e} \simeq 6 \times
10^{42}$\,erg\,s$^{-1}$, $L_{\rm disk} \simeq 3 \times 10^{46}$\,erg\,s$^{-1}$,
$L_{\rm dust} \simeq 6 \times 10^{45}$\,erg\,s$^{-1}$, $r_0 \simeq 4 \times
10^{18}$\,cm, $r \simeq 8 \times 10^{18}$\,cm, $\Gamma_{\rm j} \simeq 12$, and
$B \simeq 0.2$\,G.  For the injection electron energy distribution, the
electron Lorentz factors $\gamma_{\rm min} \simeq 1$, $\gamma_{\rm br} \simeq
1.5 \times 10^3$, and $\gamma_{\rm max} \simeq 10^6$ were obtained, along with
the spectral indices, $s_1 \simeq 0.5$ and $s_2 \simeq2.8$. The blazar model
fit to the collected dataset, and the implied physical parameters of the
4C\,+55.17 jet and its central engine, should be regarded as plausible.
Notable differences with respect to the CSO model discussed previously can be
however noted within the radio--to--X-ray frequency range.  In particular,
unlike the CSO fit, the blazar model fit does not account for the bulk of the
observed radio fluxes. These emissions, in the framework of
the blazar scenario, must therefore be produced further down the jet, at
relatively large distances from the blazar emission zone. On the other hand,
the high-energy tail of the synchrotron blazar emission dominates the radiative
output of the system around the observed near-infrared and optical frequencies, 
and also at soft X-rays. The observed hard X-ray spectrum of 4C\,+55.17 can be
hardly attributed to the IC blazar emission and requires an additional spectral
component. In general, the CSO and blazar fits differ the most within the near 
infrared and X-ray domains, hence future constraints on the hard X-ray and near 
infrared spectra, along with continued monitoring from the radio to the 
$\gamma$-ray band, should be considered as a potential way of discriminating 
between the two scenarios.

In comparing these two models, we also note the important difference between
the blazar and CSO model for 4C\,+55.17 in the radiative efficiency of the
emission zone. Compact emission zones of blazar sources are typically
characterized by a very low (less than a few percent) radiative efficiency
\citep[e.g.,][]{sik94}.  In this context, only a small fraction of the jet
kinetic power is dissipated in the blazar emission zone and radiated away in
the form of high-energy emission, which is strongly Doppler-boosted in the
observer frame due to the relativistic bulk velocity of the emitting plasma.
This is also the case for 4C\,+55.17 when modeled in the framework of the
blazar scenario discussed above.  On the other hand, the radiative efficiency
of the sub-relativistically expanding lobes of young radio sources is known to
be large, often exceeding $10\%$ \citep{young93,sta08}, which naturally
accounts for the particularly high intrinsic radio luminosity of these sources,
being comparable to the most powerful radio galaxies and quasars \citep{rea96}.
Likewise, when modeling 4C\,+55.17 as a CSO, the radiative efficiency was
similarly high.  The improved radiative efficiency of CSO sources, together
with the relatively high jet kinetic power implied by the young radio source
scenario (higher than that implied by the blazar model), can thus account for
the observed $\gamma$-ray luminosity even in the absence of relativistic
beaming.

While the CSO-type and blazar modelings of the broad-band spectrum of
4C\,+55.17 can both account for the $\gamma$-ray emission from the source, we
find the implied value for the bulk Lorentz factor $\Gamma_{\rm j} \simeq 12$
under the blazar scenario difficult to reconcile with its observed VLBI
properties. The physical mechanism responsible for the steady $\gamma$-ray
emission is also not easily explained under this framework.  Still, the unusual
characteristics of 4C\,+55.17 as for a young radio source may be evidence for a
combination of radiation produced in the sub-pc scale relativistic jet and the
emission of the compact lobes. The modeling of this complex scenario, which
might require a combination of the two models discussed above, is beyond the
scope of the present work.  A similar situation was recently considered by
Migliori (in prep), who have studied the high-energy (X-ray to $\gamma$-ray)
emission of radio-loud quasars with CSO-type inner radio morphology, such as,
e.g., 3C 186.  Objects of that type might be very common in scenarios of
intermittent jet production in active galaxies, proposed to account for the
evolution of radio-loud AGNs \citep[e.g.,][and references
therein]{rey97,sie07,cze09}.  With its complex radio structure featuring inner
and outer lobes, as well as jet-like features \citep{ros05,tav07}, 4C\,+55.17
might thus be another example of AGN with intermittent jet production.

\subsection{High Energy $\gamma$-ray Continuum of 4C\,+55.17}
\label{section-EBL}

At energies $\gtrsim 10$\,GeV the $\gamma$-ray continua of high-redshift
sources begin to suffer from substantial attenuation by the still poorly known
EBL photon field due to the photon-photon pair creation process \citep{hau01}.
By attributing the attenuation of AGN $\gamma$-ray spectra to these
interactions, it is thus possible to place significant upper limits to the EBL
provided some estimate of the source's intrinsic spectrum \citep{aha06}.  In
this respect, combined {\it Fermi} and VHE measurements by Cherenkov telescopes
such as MAGIC, H.E.S.S., and VERITAS, continue to prove successful at providing
these limits \citep[e.g.][]{geo10,ale11,orr11}.  Furthermore, with the VHE
detection of the FSRQ 3C\,279 ($z=0.536$) by MAGIC \citep{alb08}, and the
recently announced detections of others quasars -- PKS\,1510--089 ($z=0.361$)
by H.E.S.S. \citep{wag10} and PKS~1222+216 ($z=0.432$) by MAGIC \citep{ale11}
-- the search for increasingly distant luminous sources in the observable range
of ground-based Cherenkov Telescopes has become one of considerable interest to
the TeV community.  

\begin{figure}
\epsscale{0.95}
\plotone{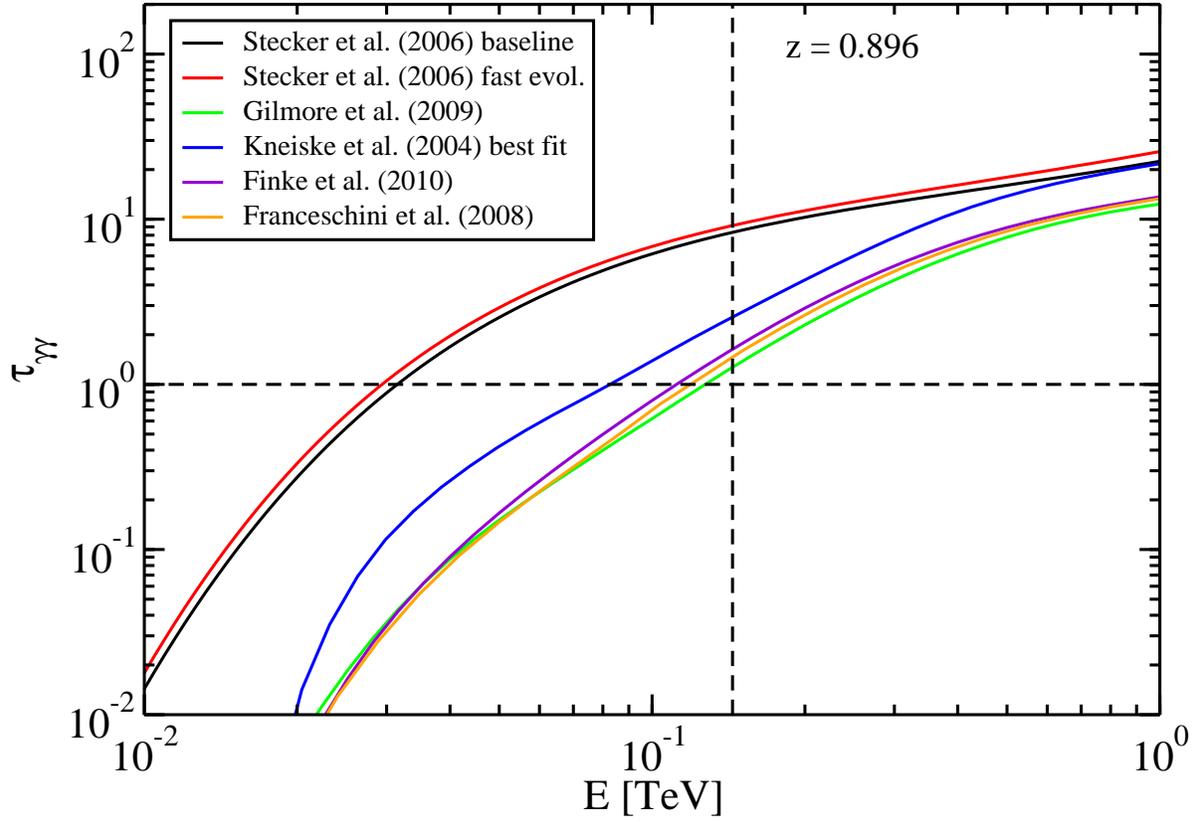}
\caption{The $\tau_{\gamma\gamma}$ opacity versus energy for several
EBL models at $z=0.896$. The highest-energy photon of 145\,GeV (rest frame
energy = 275 GeV) within the 95\%
containment radius of the 4C\,+55.17 position is also indicated (vertical
dashed line).  The horizontal line simply denotes $\tau_{\gamma\gamma}=1$.
At the observed energy, attenuation from the EBL is expected
even for those models which predict low levels of EBL.
}
\label{figure-tau}
\end{figure}

The extension of the observed $\gamma$-ray spectrum of 4C\,+55.17 up to
energies of $145$\,GeV, coupled with the source's relatively high redshift of
$z=0.896$, immediately places it among the most important high-$z$ objects that
can be used for constraining the widely debated EBL level even within LAT
energies; for an overview of different methods for constraining the EBL with
the {\it Fermi}/LAT, see \citet{ebl}. Figure\,\ref{figure-tau} illustrates the
$\tau_{\gamma\gamma}$ opacity at the redshift $z=0.896$ due to $\gamma$-ray
absorption with the EBL intensity and spectral distribution for various models
\citep{fin10,fra08,gil09,kne04,ste06} considered as a function of photon
energy. The highest-energy photon associated with 4C\,+55.17 is also indicated.
As illustrated in the figure, attenuation due to the EBL-related absorption of
$\gamma$-rays within the observed range is predicted in all the scenarios,
including those close to the lower limits derived from galaxy counts
\citep[e.g.,][]{fra08,fin10,gil09}.  

To test the validity of particular models of the EBL using the
4C\,+55.17 spectrum, we followed the likelihood ratio test method described in
\cite{ebl}. The full $>100$\,MeV observed spectrum was first fit to a broken
power law with EBL attenuation from 9 separate EBL models
\citep{fin10,fra08,gil09,pri05,ste06,sal98,kne04}, with the normalization of
the attenuation parameter $\tau_{\gamma\gamma}(E,z=0.896)$ fixed to 1 at all
energies.  The results from each of the spectral fits, including the low
($\Gamma_{\rm 1}$) and high ($\Gamma_{\rm 2}$) broken power law indices, as
well as the integral flux values, are summarized in Table\,\ref{tab:spectrum}.
Allowing the normalization of the predicted opacity $\tau_{\gamma\gamma}$ to
remain free, we then compared each result with the likelihood values obtained
when the normalization parameter was fixed to 1.  In cases where the
$\tau_{\gamma\gamma}$ normalization was reduced, a rejection at the level of
$n$ standard deviations ($\sigma$) of the particular model could be established
using the formula: 

\begin{equation}
  n=\sqrt{-2 \times \left[\log\left(L_{\rm fixed}\right)-\log\left(L_{\rm free}\right)\right]} \, ,
  \label{equation:ratio}
\end{equation}

\noindent where $L_{\rm fixed}$ and $L_{\rm free}$ are the likelihood values of
the fits for fixed and free normalizations on $\tau_{\gamma\gamma}$,
respectively. Using these results, we were able to reject two separate models
at $>3\sigma$ significance. These were the \cite{ste06} baseline and fast
evolution models at $3.9\sigma$ and $4.3\sigma$, respectively, with preferred
normalizations of $0.17\pm0.14$ and $0.16\pm0.12$.  These models were similarly
rejected in \citet{ebl} by applying the likelihood ratio test to several
blazars and gamma-ray bursts with redshifts ranging from $z=1.05$ to
$z=4.24$.  Combining this result with the overall rejection significance of the
\citet{ste06} baseline model of 11.4$\sigma$ as calculated in \citet[][\
\S\,$3.2.3$ therein]{ebl}, we obtain a new combined rejection of 11.7$\sigma$
for both the baseline and fast evolution models.

\begin{figure}
\epsscale{0.95}
\plotone{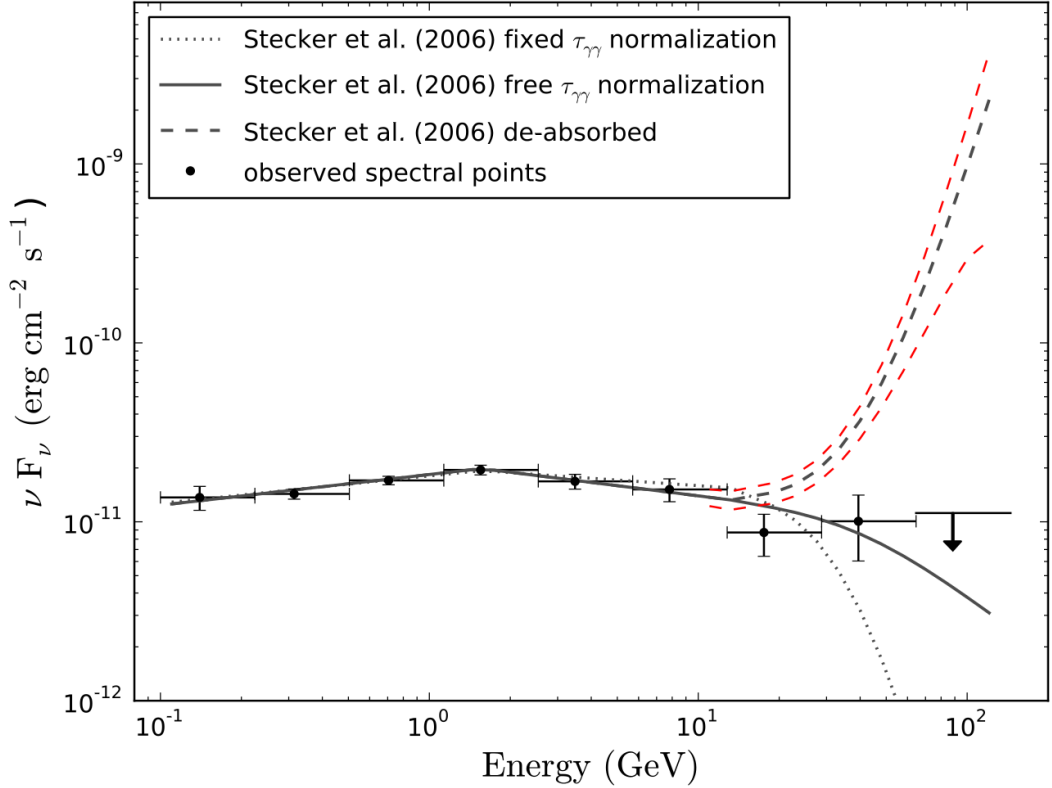}
\caption{The de-absorbed spectrum of 4C\,+55.17 (thick dashed line, gray) along with 1$\sigma$ error
bars (thin dashed lines, red) using the \citet{ste06} baseline model.  Observed spectral points
without de-absorption, along with the observed spectrum with
$\tau_{\gamma\gamma}$ normalization left free (solid line) and fixed to 1
(dotted line), are plotted for comparison.  The de-absorbed spectrum shows the
non-physical behavior of an unbounded exponential rise up to the observed LAT
energy of 145 GeV.  This trend, which is preferred by $3.9\sigma$ over a single
power law, increases the intrinsic spectrum by two orders of magnitude above
the observed inverse Compton peak and requires the modeling of an additional
(and unknown) spectral component (see Figures \ref{figure-model} and 
\ref{figure-model-bl}). 
}
\label{figure-stecker}
\end{figure}

Figure\,\ref{figure-stecker} shows the predicted shape of the intrinsic
spectrum of 4C\,+55.17 obtained by de-absorbing the observed {\it Fermi}
spectrum using the \citet{ste06} baseline EBL model.  A common feature
occurring from models which over-predict the level of EBL is that of an
unbounded exponential spectral rise at highest energies -- a behavior which can
largely be considered non-physical, and has thus been used in previous studies
to place constraints on the EBL using TeV observations \citep[e.g.,][]{dwe05}.
This behavior is clearly illustrated in the case of the \citet{ste06} baseline
model.  Such a feature would in turn require the modeling of an additional
spectral component beyond that which we consider in \S\,\ref{section-model} and
that would be orders of magnitude more luminous than the observed
inverse-Compton (IC) peak.  We also note that any intrinsic $absorption$ that
may be taking place within the source represents an even greater rejection of
this model, as the true attenuation due to the EBL would be even less.  We
therefore consider the \citet{ste06} baseline and fast
evolution\footnote{Because the \citet{ste06} fast evolution model predicts an
increased opacity from the baseline model, our conclusions from the baseline
test can be applied in both cases.} models to over-predict the true level of
EBL at the observed redshift and energies.  

\begin{figure}
\epsscale{1.1}
\centering
\plotone{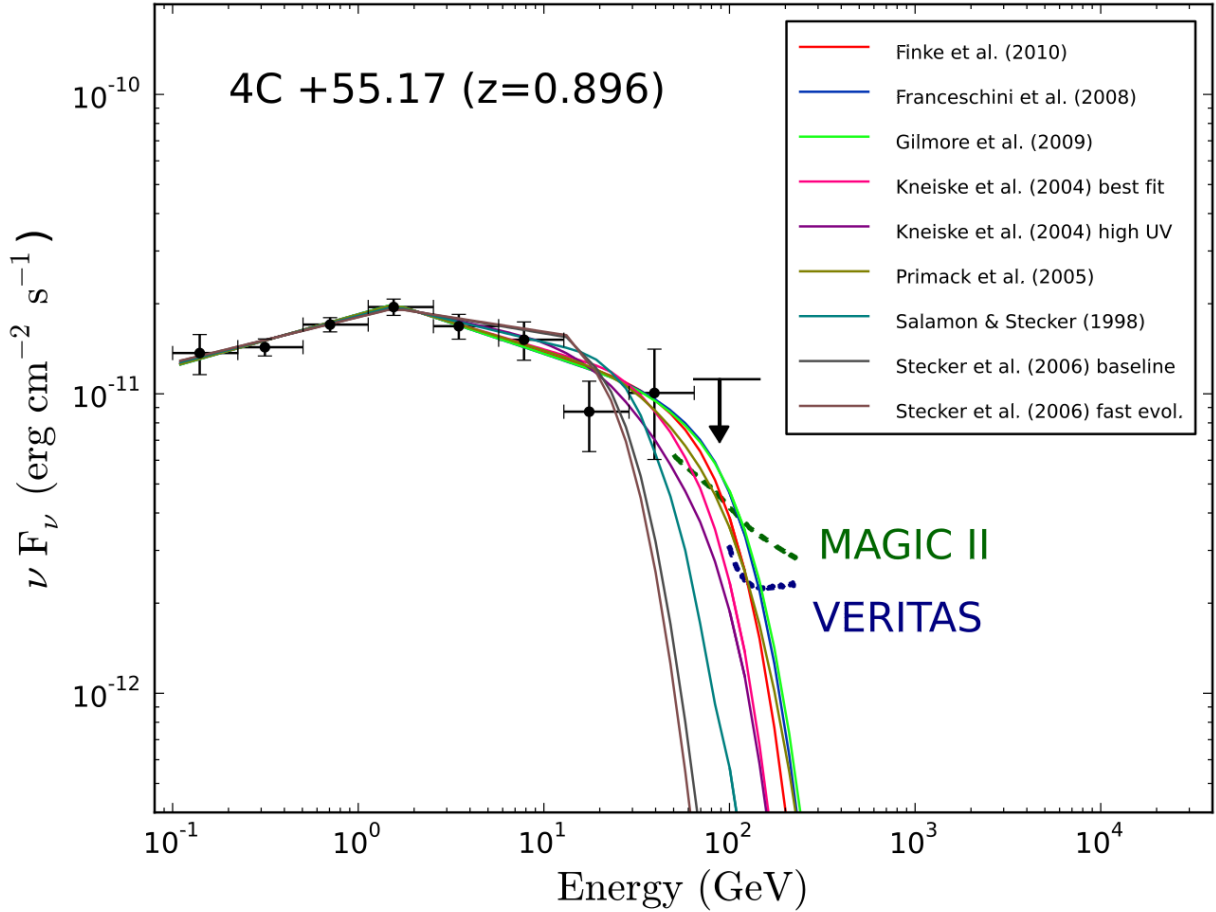}
 \caption{The observed LAT spectrum fit to a broken power law with attenuation
 from 9 different EBL models.  The spectra are extrapolated beyond the observed
 energy of 145 GeV and compared against the MAGIC\,II and VERITAS differential
 flux sensitivity curves for a 50 hour, 5$\sigma$ detection of a source
 characterized by an exponentially decreasing spectrum (see 
 Appendix\ \ref{section-sensitivity}).  For several EBL models, the
 4C\,+55.17 spectrum is found to intercept with both the VERITAS and MAGIC\,II
 sensitivities, making 4C\,+55.17 a viable candidate for future ground-based
 VHE observations.  
}
\label{figure-sed}
\end{figure}

With its excellent sensitivity in the high-energy range, the LAT instrument
provides a unique opportunity to search for VHE candidates at high redshifts
through detailed spectral analysis of the {\it Fermi} data.  In the case of
4C\,+55.17, the attenuated high-energy spectrum obtained from fitting the nine
often discussed EBL models is illustrated in Figure~\ref{figure-sed}.  Each
spectrum is extrapolated beyond the highest observed photon energy of
$145$\,GeV and compared against the typical differential flux sensitivity
curves of currently operating TeV telescopes.  With the exception of the four
``highest-level'' EBL models (including the two models ruled out by the present
work), the observed 4C\,+55.17 spectrum is found to lie at the observable
threshold for ground-based observations. It is also worth noting that while
intrinsic absorption from interactions with the UV disk and infrared torus may
contribute to the spectral attenuation at energies $>100$\,GeV, this effect
would be reduced in cases where the $\gamma$-ray emission takes place at
hundreds-of-parsecs scale distances from the central black hole, for which
there is compelling evidence in the case of 4C\,+55.17 (see
\S\,\ref{section-cso}).  In addition, with the present analysis we find no
evidence of variability in 4C\,+55.17 over 19 months of LAT observing time, and
furthermore we find its flux to be consistent with the EGRET measured value,
thus showing no evidence of variability at $\gamma$-ray energies over decade
timescales as well.  The non-variable $\gamma$-ray continuum of the source thus
promises the opportunity to observe it over the extended timescales required
for a $5\sigma$ detection.  This is in contrast to the other VHE-detected
quasars, which were detected only during periods where the sources were in a
flaring state.  In this way 4C\,+55.17 stands apart from all of the
EBL-constraining sources considered in \citet{ebl}, as it holds the greatest
potential for providing future constraints.

\section{Conclusions}\label{section-conclusions}

The investigation of the multiwavelength properties of 4C\,+55.17, including
its unusually hard $\gamma$-ray spectrum, lack of distinct variability, and
CSO-like radio morphology, has highlighted the exceptional nature of this
$\gamma$-ray source.  For the first time, we have modeled the radio to
$\gamma$-ray emission of 4C\,+55.17 as a young radio source using a dynamic
model that is consistent with the full extent of its observed properties.
Furthermore, we find that the prospect of a VHE observation of 4C\,+55.17,
whose $\gamma$-ray spectrum already extends up to the observed energy of 145\
GeV, is within reach of the current generation of Cherenkov telescopes.  A
detection by such an instrument would not only add to the present understanding
of the source itself, but would also serve to place a significant upper limit
to the level of EBL through combined {\it Fermi} and VHE data.  Furthermore, we
anticipate that through continued monitoring of 4C\,+55.17 at high energies
with the {\it Fermi} LAT, as well as in the radio through X-rays, the precise
classification of 4C\,+55.17 will become increasingly more apparent.  If, for
example, the source continues to remain non-variable in $\gamma$-rays in the
years to come, its average flux versus variability may eventually lie outside
the distribution of {\it Fermi} $\gamma$-ray emitting blazars altogether, which
would make any standard blazar emission scenario difficult to reconcile.  On
the other hand, if rapid variability is found in this source, that would seem
to rule out a pure CSO interpretation. Thus we expect that 4C\,+55.17 will be
an important target for future observations across all wavelengths.

\acknowledgments

The \textit{Fermi} LAT Collaboration acknowledges generous ongoing support
from a number of agencies and institutes that have supported both the
development and the operation of the LAT as well as scientific data analysis.
These include the National Aeronautics and Space Administration and the
Department of Energy in the United States, the Commissariat \`a l'Energie Atomique
and the Centre National de la Recherche Scientifique / Institut National de Physique
Nucl\'eaire et de Physique des Particules in France, the Agenzia Spaziale Italiana
and the Istituto Nazionale di Fisica Nucleare in Italy, the Ministry of Education,
Culture, Sports, Science and Technology (MEXT), High Energy Accelerator Research
Organization (KEK) and Japan Aerospace Exploration Agency (JAXA) in Japan, and
the K.~A.~Wallenberg Foundation, the Swedish Research Council and the
Swedish National Space Board in Sweden.

Additional support for science analysis during the operations phase is gratefully
acknowledged from the Istituto Nazionale di Astrofisica in Italy and the Centre National d'\'Etudes Spatiales in France.

L.O. acknowledges support by a 2009 National Fellowship ``L'OR\'EAL Italia
Per le Donne e la Scienza'' of the L'OR\'EAL-UNESCO program ``For Women in
Science,'' and partial support from the INFN grant PD51 and the ASI Contract
No.\ I/016/07/0 COFIS.
{\L}.S. is grateful for the support from Polish MNiSW through the grant N-N203-380336.
R.M. was supported by the MNiSW grant no. N-N203-301635

The authors acknowledge the support by the $Swift$ team for providing ToO observations
and the use of the public HEASARC software packages.

The authors would like to thank Annalisa Celotti, Luigi Costamante, Berrie Giebels,
and Dave Thompson for their helpful comments and suggestions.  

\appendix
\section{Association of the 145 GeV photon with 4C\,+55.17}
\label{section-photon}

To further investigate the VHE detection of the source, the $145$\,GeV event
was analyzed in detail using the event
display\footnote{\url{http://glast-ground.slac.stanford.edu/DataPortalWired/}}
and found to be a clean $\gamma$-ray event, going through more than half a
tracker tower before interacting in the back planes and generating a
well-behaved symmetric shower in the calorimeter. A full Monte Carlo simulation
was also run in order to determine the accuracy of the energy reconstruction. A
total of $500,000$ $\gamma$-rays between the energies $50$ and $200$\,GeV were
simulated at an incoming angle $\theta$ and $\phi$ equivalent to that of the
measured event. Data selection cuts were applied on all similar variables,
including cuts on the calorimeter raw energy, best measured energy,
reconstructed direction, and event class level. The distribution in Monte Carlo
energy for the remaining events was found to give a $\sim1\sigma$ error of $\pm
11$\,GeV.

The probability of the $145$\,GeV event occurring by random coincidence from
background contamination was calculated using the \texttt{gtsrcprob} analysis
tool. Probabilities of each event are assigned via standard likelihood analysis
to all sources within a provided best-fit model \citep{mat96}. The probability
that a photon is produced by a source $i$ is proportional to $M_{\rm i}$, given by
the formula: 

\begin{equation} 
  M_{\rm i}(\varepsilon',\hat{p}',t)=\int_{\rm SR}
  d\varepsilon \, d\hat{p} \,\, S_{\rm i}(\varepsilon,\hat{p}) \,\,
  R(\varepsilon',\hat{p}';\varepsilon,\hat{p},t) \, , 
  \label{eqn:prob}
\end{equation}

\noindent where $S_{\rm i}(\varepsilon,\hat{p})$ is the predicted counts
density from the source at energy $\varepsilon$ and position $\hat{p}$, and
$R(\varepsilon',\hat{p}';\varepsilon,\hat{p},t)$ is the convolution over the
instrument response. In this way, all the surrounding point sources, the
diffuse background, and their corresponding best-fit spectra are taken into
account when assigning probabilities to individual photon events. For the
$145$\,GeV event, the probability of spurious association with 4C\,+55.17 was
found to be $1.8\times10^{-3}$, agreeing well with an independent method by
\cite{ner10}, who quote a chance probability by background contamination of
$3.1\times10^{-3}$ for the same event. 

\section{Calculation of the MAGIC\,II and VERITAS Differential Flux Sensitivities}
\label{section-sensitivity}

Starting with the integral flux sensitivity curves of MAGIC\,II \citep{bor10} and
VERITAS \citep{per09}, the differential flux sensitivities can be derived for a
given functional form.  In the case of 4C\,+55.17, we may represent the
attenuated VHE spectrum in general with an exponential cutoff given by the
formula:

\begin{equation} 
  \frac{dN}{dE} = N_{\rm 0} \, E^{-\Gamma} \, e^{-(\frac{E}{E_{\rm c}})} 
  \label{eqn:expo_cutoff}
\end{equation}

\noindent where $N_{\rm 0}$, $E_{\rm c}$, and $\Gamma$ are free parameters of
the fitted form of the function.  The integral flux above some minimum
energy $E_{\rm o}$ is thus given by:

\begin{equation} 
  N = N_{\rm 0} \, \int_{E_{\rm 0}}^{\infty}
  dE \,\, E^{-\Gamma} \,\, e^{-\frac{E}{E_{\rm c}}}
  \label{eqn:dNdE_integrated}
\end{equation}

\noindent Defining the quantity

\begin{equation} 
  \Psi(E) \equiv \int_{E_{\rm 0}}^{\infty}
  dE \,\, E^{-\Gamma} \,\, e^{-\frac{E}{E_{\rm c}}}
  \label{eqn:Psi}
\end{equation}

\noindent the appropriate solution for $N_{\rm 0}$ may be substituted into
equation \ref{eqn:expo_cutoff} to obtain:

\begin{equation} 
  \frac{dN}{dE}\Bigg |_{E_{0}} = \frac{N  \, E_0^{-\Gamma} \, e^{-(\frac{E_0}{E_{\rm c}})}}{\Psi(E_0)} 
  \label{eqn:dNdE}
\end{equation}

\noindent To construct the differential flux sensitivity curves, we obtained the values
$\Gamma=2.12$ and $E_{\rm c}=100$ GeV by performing a \texttt{gtlike} fit of
the $>1.6$ GeV data of 4C\,+55.17 to the exponential cutoff functional form.
For each value $N$ of the integral flux sensitivity, a corresponding
differential flux sensitivity value could thus be obtained via numerical
evaluation of equation \ref{eqn:dNdE}.

{}

\clearpage

\begin{deluxetable}{lcccc}
\tabletypesize{\scriptsize}
\tablecaption{{\it Swift}/UVOT Observations of 4C\,+55.17}
\tablewidth{0pt}
\tablehead{
\colhead{Band} & \colhead{$\lambda$ [\Angst]} & \colhead{$F_{\rm ep1}$ [mJy]} & \colhead{$F_{\rm ep2}$ [mJy]} & \colhead{$F_{\rm ep3}$ [mJy]}
}
\startdata
    V & 5402 & $0.331 \pm 0.061$ & $0.337 \pm 0.029$ & .. \\
    B & 4329 & $0.262 \pm 0.015$ & $0.286 \pm 0.015$ & .. \\
    U & 3501 & $0.249 \pm 0.010$ & $0.251 \pm 0.011$ & .. \\
    UVW1 & 2634 & $0.175 \pm 0.007$ & $0.174 \pm 0.007$ & .. \\
    UVM2 & 2231 & $0.142 \pm 0.029$ & $0.167 \pm 0.007$ & .. \\
    UVW2 & 2030 & $0.125 \pm 0.009$ & $0.127 \pm 0.005$ & $0.130 \pm 0.005$  \\
\enddata
\tablecomments{The observations were obtained on 2009 March 5 (ep1), Nov 11 (ep2), and Nov 26 (ep3).}
\label{tab:uvot}
\end{deluxetable}

\begin{deluxetable}{llllllll}
\tabletypesize{\footnotesize}
\tablecaption{De-absorption of $\gamma$-ray flux using different EBL models with fixed $\tau_{\gamma\gamma}$ normalization}
\tablewidth{0pt}
\tablehead{
\colhead{EBL model} & \colhead{$\Gamma_{\rm 1}$} & \colhead{$\Gamma_{\rm 2}$} & \colhead{Flux$^a$} & \colhead{-log(likelihood)}
}
\startdata
    Finke et al. (2010) & $1.83 \pm 0.05$ & $2.20 \pm 0.06$ & $9.05 \pm 0.46$ & 595671.252 \\
    Franceschini et al. (2008)& $1.83 \pm 0.05$ & $2.21 \pm 0.06$ & $9.04 \pm 0.46$ & 595671.192 \\
    Gilmore et al. (2009)& $1.83 \pm 0.05$ & $2.21 \pm 0.06$ & $9.04 \pm 0.46$ & 595671.133 \\
    Primack et al. (2005) & $1.83 \pm 0.05$ & $2.19 \pm 0.06$ & $9.06 \pm 0.46$ & 595671.074 \\
    Kneiske (2004) best fit& $1.83 \pm 0.05$ & $2.18 \pm 0.06$ & $9.06 \pm 0.46$ & 595671.577 \\
    Kneiske (2004) high UV & $1.84 \pm 0.05$ & $2.14 \pm 0.06$ & $9.10 \pm 0.46$ & 595672.025 \\
    Stecker (2006) baseline & $1.85 \pm 0.05$ & $2.10 \pm 0.06$ & $9.14 \pm 0.46$ & 595678.519 \\
    Stecker (2006) fast evolution& $1.85 \pm 0.05$ & $2.10 \pm 0.07$ & $9.14 \pm 0.46$ & 595680.170 \\
    Salamon \& Stecker (1998) & $1.84 \pm 0.05$ & $2.15 \pm 0.06$ & $9.10 \pm 0.46$ & 595673.291 \\
\enddata
\tablenotetext{a}{Flux above $100$\,MeV in units of [$10^{-8}$\,cm$^{-2}$\,s$^{-1}$]}
 \label{tab:spectrum}
\end{deluxetable}

\clearpage

\clearpage

\clearpage

\clearpage

\clearpage

\clearpage

\clearpage

\end{document}